\documentclass[10pt]{article}
\usepackage[margin=1in]{geometry}

\usepackage{times}
\usepackage[pdftex]{graphicx}
\graphicspath{{images/}}

\usepackage{amsmath}
\usepackage{latexsym}
\usepackage{amsfonts}
\usepackage{amssymb}

\usepackage{siunitx} 
\DeclareSIUnit{\feet}{ft}

\usepackage{pdflscape} 

\usepackage{paralist}	
\usepackage{enumitem}

\usepackage[misc]{ifsym} 
 
\usepackage{xspace}

\newcommand{\eg}{e.g.,\xspace}
\newcommand{\ie}{i.e.,\xspace}

\newcommand{\etc}{etc.\xspace}

\newcommand{\optional}[1]{}

\usepackage{acronym}

\acrodef{ai}[AI]{Artificial Intelligence}
\acrodef{conops}[CONOPS]{concept of operations}
\acrodef{od}[OD]{Operational Domain}
\acrodef{odd}[ODD]{Operational Design Domain}
\acrodef{sodd}[SODD]{System Operational Design Domain}
\acrodef{ml}[ML]{Machine Learning}
\acrodef{mlc}[MLC]{ML Constituent} 
\acrodef{mlm}[MLM]{ML Model}
\acrodef{nhtsa}[NHTSA]{National Highway Traffic Safety Administration}
\acrodef{ads}[ADS]{automated driving system}
\acrodef{cav}[CAV]{connected and automated vehicle}
\acrodef{asam}[ASAM]{Association for Standardization of Automation and Measuring Systems}
\acrodef{dsl}[DSL]{domain-specific language}

\usepackage{hyperref}
\hypersetup{
    colorlinks=true,	
    linkcolor=black,	
    citecolor=black,	
    filecolor=black,    
    urlcolor=blue		
}

\usepackage{fancyhdr}
\fancypagestyle{firstpage}
{
    \fancyhead[C]{\textcolor{red}{The content of this author pre-print version has been accepted for publication after peer review but is not the Version of Record. The Version of Record appears in the Proceedings of the 42nd International Conference on Computer Safety, Reliability, and Security (SAFECOMP 2023), Toulouse, France.}}   
    }

\begin{document}

\title{Data-centric Operational Design Domain Characterization for 
		Machine Learning-based Aeronautical Products}


%
%
%

\author{Fateh Kaakai\\
			\normalsize{Thales, 3 Avenue Charles Lindbergh, 94628 Rungis, France}\\
			\normalsize{fateh.kaakai@thalesgroup.com}
	\and
				Sridhar (``Shreeder'') Adibhatla\\
			\normalsize{Rockdale Systems LLC, Cincinnati OH 45246, USA}\\
			\normalsize{adiabatic@gmail.com}
	\and
				Ganesh Pai\thanks{Contribution to the paper with support from the System-wide Safety project under the Airspace Operations and Safety Program of the NASA Aeronautics Research Mission Directorate.}\\
			\normalsize{KBR / NASA Ames Research Center, Moffett Field, CA 94401,
			 USA}\\
			\normalsize{ganesh.pai@nasa.gov}
	\and
				Emmanuelle Escorihuela\\
			\normalsize{Airbus Operations (SAS), 316 Route de Bayonne, 31060 
				Toulouse, France}\\
			\normalsize{emmanuelle.escorihuela@airbus.com}
		}

\date{}

\maketitle

\thispagestyle{firstpage}
\begin{abstract}

We give a first rigorous characterization of \acfp{odd} for \acf{ml}-based aeronautical products. Unlike in other application sectors (such as self-driving road vehicles) where \ac{odd} development is scenario-based, our approach is \emph{data-centric}: we propose the dimensions along which the parameters that define an \ac{odd} can be explicitly captured, together with a categorization of the data that \ac{ml}-based applications can encounter in operation, whilst identifying their system-level relevance and impact. Specifically, we discuss how those data categories are useful to determine:  the requirements necessary to drive the design of \acp{mlm}; the potential effects on \acp{mlm} and higher levels of the system hierarchy; the learning assurance processes that may be needed, and system architectural considerations. We illustrate the underlying concepts with an example of an aircraft flight envelope. 
\optional{The approach in this paper is one of the cornerstones of a future process guidance for development and certification/approval of safety-related aeronautical products implementing \ac{ai}, currently being developed through aviation industry-based consensus, jointly by the SAE G-34 Committee for \ac{ai} in aviation, and EUROCAE WG-114 for \ac{ai}.}

\end{abstract}

\section{Introduction}\label{s:introduction}

\acf{ai}-enabling technologies like \acf{ml} have the potential to transform the aviation industry by creating new products and services, and by enhancing the existing ones. However, \ac{ml} introduces a new paradigm for design activities since the intended behavior of a function is inferred from a body of data using statistical learning algorithms, rather than being specified and programmed. Data is thus central to the implementation of a final product design.  

In traditional aviation domain systems engineering, operational requirements capture the conditions under which an end-product is expected to fulfill its missions. Those requirements, which are an expression of stakeholder needs, contain parameters and values that define an operational environment, or \emph{operational domain} (\acs{od}), in which an aviation system must properly operate. When requirements are elicited from and allocated to different layers of the system design---namely: \emph{function} or \emph{system}, \emph{subsystem}, and eventually \emph{item}\footnote{We use the standard aviation domain terminology for the layers of a system hierarchy/design.}---the \acs{od} is also correspondingly allocated, resulting in \emph{operational design domains} (\acsp{odd}) corresponding to those layers.

\acused{odd}
\acused{od}

\subsection{Motivation and Contributions}\label{sss:motivation-contributions}

Specifying, refining, and allocating \acp{od} to the system layers that will eventually integrate \ac{ml} are activities not as well-understood as they are when going from a system/function layer to lower layers in conventional aviation systems development processes.
Thus, a key challenge for the aviation systems domain is how to define, analyze, and manage the \acp{odd} resulting from the allocation of the \ac{od} to the system layers integrating \ac{ml}.\footnote{Additional related challenges (not in scope for this paper), such as the need to adapt requirements definition and validation processes to account for dataset requirements, have been comprehensively elaborated in~\cite{SAE-AIR-6988}.} Addressing this challenge is especially important because it not only drives the data collection activities needed to ensure that a dataset representative of the intended operations is gathered, but also it influences the design of those layers and the underlying \acfp{mlm}. 

To that end, this paper makes the following main contributions: 
\begin{inparaenum}[(i)]
	\item in Section~\ref{sss:odd-parameters}, an identification of the dimensions along which the parameters that define an \ac{odd} for \ac{ml}-based aeronautical products can be explicitly captured;  
	\item in Section~\ref{s:concepts}, a rigorous data-centric characterization of \acp{odd} based on categorizing the data that \ac{ml}-based functionality can encounter in operation. An aircraft flight envelope example also concretizes the underlying concepts; and
	\item in Section~\ref{s:odd-impact}, an illustration of how the identified data categories can be used to determine the potential effects on the system layer integrating \ac{ml}, along with the learning assurance activities and the system architectural considerations needed to mitigate those effects. 
\end{inparaenum}

The approach in this paper is one of the cornerstones of a future process guidance document~\cite{as6983-draft} for the development and certification/approval of safety-related aeronautical products implementing \ac{ai}. That guidance is currently being developed through an aviation industry-based consensus process, jointly by the SAE Committee for \ac{ai} in Aviation (G-34), and the EUROCAE working group for \ac{ai} (WG-114).

\subsection{Related Work}\label{sss:related-work}

The concept of \ac{odd} was initially introduced and developed by the automotive systems industry~\cite{sae-j3016}. As such, the current literature on specifying, developing, and using \acp{odd} is largely in an automotive systems application context. 
For example, \ac{odd} specification for \acp{ads} can be aided by a \ac{dsl} using structured natural language founded on a formal, machine-processable domain model, to support both human comprehension and programmatic manipulation~\cite{odd-definition-language}. A \emph{divide-and-conquer} approach to automotive function \ac{odd} development can be employed using a concept of so-called $\mu$ODD~\cite{micro-odd} that partitions an \ac{odd} to place useful bounds on various safety-relevant parameters. Such partitions can then be tied to validation tests, whilst also encoding situation-specific parameter information. This approach is closest to our work, although the partitioning we achieve is data-centric, and orthogonal to $\mu$ODD-based partitions. 
In~\cite{ads-tests-scenarios}, a hierarchical \ac{odd} definition is used to develop a scenario-based test framework for \acp{ads}. 

The \ac{odd} concept is being progressively matured in the automotive industry via \ac{odd}-related standards~\cite{bsi-odd-pas}, \cite{iso-dis-34503}, as well as automotive system-centric safety standards concerning \ac{ml} and \ac{ai}~\cite{iso-pas-8800}, \cite{ul4600std}. Each of those guidance documents gives a mutually consistent definition for the \ac{odd} concept, emphasizing its relationship to safety. Nevertheless, automotive domain guidance cannot be directly applied to safety-critical aeronautical products owing to a variety of constraints, including: 
\begin{inparaenum}[(a)]
	\item differences in the regulatory approach between the automotive and aviation sectors; 
	\item the need for standards to be compatible with aviation regulations and regulatory acceptance of the associated compatibility arguments; 
	\item the stringency of assurance requirements in the aviation sector; and 
	\item consistency with the existing ecosystem of recommended engineering practices, \eg for safety assessment~\cite{SAE-ARP-4761}, and aviation system development~\cite{SAE-ARP-4754A}.
\end{inparaenum}

All of those factors, besides the key challenge discussed earlier, have additionally motivated the work in this paper. Next we give our notion of \ac{odd}.

\section{System-level Considerations}\label{s:system-level-considerations} 

\subsection{\acfp{od}} 

When designing a product system, it is an established and well-understood aviation systems engineering practice to capture and analyze stakeholder needs at an early stage, along numerous dimensions such as the mission to be fulfilled, the expected performance in different system operating phases, and specific environmental conditions encountered. 
An \ac{od} is one of the results of such early-stage analysis, and it is embodied by the operational requirements for that system. In other words, the \ac{od} is captured in the form of requirements via a \emph{specification} activity of a well-defined requirements development process. 
Thus, we consider an \ac{od} to be a specification of all foreseeable operating conditions under which an end-product is expected (and should be designed) to fulfill its missions. For instance, a \emph{flight envelope} specifies, at a minimum, a combination of altitude and \emph{Mach number}\footnote{Mach number is the ratio of true airspeed to the local speed of sound.} values that define an operational environment in which an aircraft type must properly operate.

\subsection{\acfp{odd}}

We define the \emph{allocation of an \ac{od}} to be the \emph{operational design domain} (\ac{odd}). This is largely aligned with other definitions of \ac{odd}~\cite{nhtsa-safety-vision}, \cite{as6983-draft}, \cite{sae-j3016}, \cite{ul4600std}. 
Just as requirements are allocated across the different layers of the system design, and then refined with various criteria in mind, \eg safety, architectural options, implementation choices, and physical considerations, an \ac{od} is also allocated to the lower design layers, and further refined so that each layer has its own \ac{odd}, \ie the portion of the associated \ac{od} in which it should properly function. Such refinement can potentially (but not always) lead to rich and complex \acp{odd}.\footnote{Characterizing the complexity of an \ac{odd} is not in scope for this paper.}
The principles and procedures governing \ac{od} allocation rely upon established aerospace practices~\cite{SAE-ARP-4754A}. 
As such, we can allocate the entire \ac{od} or a portion thereof to the subsystems that will be implemented using \ac{ml} technologies (which we refer to, henceforth, as \emph{ML-based subsystems}). Moreover, refining requirements as indicated earlier will bring forth corresponding enhancements of the \ac{od} reflecting the same considerations. 
\optional{For example, an \ac{od} at a system layer could initially contain details about airports, runways, and air routes, say. By the time it is refined and eventually allocated to the lower system layers, the resulting \ac{odd} would contain additional details, such as the vibration levels, blur tolerances for images captured by optical sensors, together with information about sensor mounting configurations and orientations, and so on.}

\subsection{Describing \acp{od} and \acp{odd}}\label{sss:odd-parameters}

To describe an \ac{od} or \ac{odd} we elicit a variety of \emph{parameters}, their range of admissible values, and, when relevant, distributions of occurrences over particular time intervals. In general, these define a multi-dimensional region. In practice, an \ac{od} or \ac{odd} is often likely to be a subset of that region. Although there are many ways to group parameters, the following is typical in practice: 
\begin{compactitem}
	\item \emph{Environmental Parameters}: These are variables outside the product (\eg aircraft) system boundary, including weather conditions (ambient air temperature and pressure, wind conditions, humidity/rain/snow/ice, dust or sand levels, \etc) as well as application-specific parameters, \eg brightness, contrast levels, and blur levels for optical sensor systems.
  
  \item \emph{Operational Parameters}: These are parameters within the system boundary, examples of which include altitude and Mach number limits specified by a flight envelope, as well as ranges for angle of attack, pitch, roll, yaw angles, or their rates of change. 
	
	\item \emph{System Health Parameters}: These specify whether the system is expected to work only under nominal (non-failure) conditions, or whether it should be able to handle deterioration over time, sensor failures, or failures in specified system components (\eg a failed actuator or a damaged flight control surface).
\end{compactitem}

\subsection{\ac{mlc}}

Traditional systems engineering activities need to transition to \ac{ml} activities at a certain stage of system development when integrating \ac{ml}. In light of this, current regulatory guidance for introducing \ac{ml} technologies into safety-related aeronautical applications~\cite{easa-concept-paper}, as well as ongoing standardization activities~\cite{as6983-draft} have introduced a concept of \emph{\acl{mlc}}~(\acs{mlc}) for systems integration purposes. 

Effectively, an \ac{mlc} represents the lowest-level of a functional decomposition from a system perspective that supports a subsystem function. It is a grouping of hardware and/or software items implementing one or more \acfp{mlm} and their associated data pre- and post-processing items. Pre-processing may include (but is not restricted to) data cleanup, normalization, and feature computation. Similarly, post-processing may involve, among other actions, denormalization and blending of outputs from sub-models. 

We qualify the \ac{odd} based on its allocation. Thus, allocating an \ac{od} to an \ac{mlc} gives an MLCODD (\ie the design space for an \ac{mlc}), and likewise, the allocation to an \ac{mlm} results in an MLMODD. An MLMODD may be identical to the MLCODD, though in practice it may be smaller. Additionally, an \ac{mlc} can contain multiple \acp{mlm} each of which have their respective MLMODDs.
Also, an MLMODD (or MLCODD) may be the same as the \ac{od} for the system, its superset (to provide robustness), or a subset thereof (to limit the design to a feasible region). 
Thus, when a product will eventually integrate \ac{ml} (\eg as software whose design was learned through an \ac{ml} training process) understanding the MLCODD is crucial to ensure that:
\begin{inparaenum}[(1)]
	\item the data used for training is representative of that \ac{od}; and 
	\item the \ac{ml} designer comprehends the complexity of the portion of the \ac{od} that has been allocated to machine learned functionality.
\end{inparaenum}

\section{New \ac{odd} Concepts for Aviation}\label{s:concepts}

From the preceding narrative, it should be evident that developing an \ac{od}/\ac{odd} is itself not a new phenomenon in aviation systems engineering practice. However, it is the transition from an \ac{od}/\ac{odd} description to data collected for \ac{mlm} training that is the major change relative to the way \acp{od} are typically specified during conventional (\ie non-\ac{ml} based) product development. This change requires alternative approaches that are the focus of learning assurance processes~\cite{easa-concept-paper}, \cite{mldl-sae-journal}. 

We now give a data-centric conceptual characterization for \acp{odd}, that partitions them based on \emph{categories} and \emph{kinds} of data. Henceforth, when we refer to ``\ac{odd}'' and ``\ac{ml}'', we mean the MLMODD (or MLCODD), and the MLM (or MLC), respectively, and we will qualify our usage of those terms when it is not clear from context.

\subsection{Categories and Kinds of Data}\label{sss:data-categories}

We define the following data categories: 

\begin{compactenum}[(i)]
	\item \emph{Nominal}: Set of data points that lie in the interior of an \ac{odd} statistical distribution, that is \emph{correct} with respect to the corresponding ML requirements.
	
	\item \emph{Outlier}: Set of data points outside an \ac{odd}. Some data can be mistaken to be \emph{Outlier} data when they should have been \emph{Nominal} data, had that \ac{odd} been correctly characterized by including at least one additional parameter.

	\item \emph{Edge Case}: Set of data points on an \ac{odd} boundary where exactly one \ac{odd} parameter has a valid extreme (maximum and minimum) value.
 
	\item \emph{Corner Case}: Set of data points where at least one \ac{odd} parameter is at their respective extremum (minimum and maximum value) of the range of values for those parameters that are admissible (or valid) for a given \ac{odd} (see Figure~\ref{f:odd-flight-envelope-example} for examples). There are two types of \emph{Corner Case} data:
\begin{compactitem}
	\item \emph{Feasible}: those that are part of the functional intent and, thus, within a given \ac{odd} (specifically at the vertices\footnote{\acp{odd} without vertices \eg an oval region, will therefore not have feasible corner cases.} of that \ac{odd}); 
	\item \emph{Infeasible}: those that are not part of the functional intent and, thus, outside the \ac{odd}. Note that all \emph{Infeasible Corner Case} data are a special case of \emph{Outlier} data.
\end{compactitem}

	\item \emph{Inlier} $(\mathtt{InL})$: Set of data that lie in the interior of the ODD following an error during data management, \eg due to incorrect usage of units and dimensions. \emph{Inlier} data are difficult to distinguish from \emph{Nominal} data, and hence difficult to detect/correct. 
	
	\item \emph{Novelty}: Set of data within an \ac{odd} according to the parameters used to describe that \ac{odd}, but which should have been considered to be \emph{Outlier} data, had that \ac{odd} been correctly described by introducing at least one additional \ac{odd} parameter. In this sense, \emph{Novelty} data points for an \ac{odd} could be seen as \emph{duals} of those data points that are mistakenly considered to be \emph{Outlier} data, when they should, in fact, have been \emph{Nominal} data for that \ac{odd}. \emph{Novelty} data usually arise from insufficient \ac{odd} characterization. 
		
\end{compactenum}
\optional{In general---especially when considering \emph{Novelty} and \emph{Outlier} data---we can update an \ac{odd} to include any missing parameters according to established update rules.}

We can group the \emph{Inlier}, \emph{Outlier} (including \emph{Infeasible Corner Case}), and \emph{Novelty} categories into a single \emph{Anomaly} data category. 
Data drawn from all the aforementioned categories may also be characterized as among the following kinds of sets:


\newcommand{\inmlmodd}{\mathtt{InMOD}}
\newcommand{\outmlmodd}{\mathtt{OutMOD}}
\newcommand{\inmlcodd}{\mathtt{InCOD}}
\newcommand{\outmlcodd}{\mathtt{OutCOD}}

\newcommand{\insample}{\mathtt{InS}}
\newcommand{\outofsample}{\mathtt{OutS}}

\newcommand{\inlier}{\mathtt{InL}}
\newcommand{\outlier}{\mathtt{OutL}}
\newcommand{\rdo}{\mathtt{RDO}}

\begin{compactenum}[(a)]
	\item \emph{In-Sample} $(\insample)$: Data used during \ac{mlm} learning which the implementation of the \ac{mlm} will have to process during inference in operation.

	\item \emph{Out-of-Sample} $(\outofsample)$: Data not used during \ac{mlm} learning that the implementation of the \ac{mlm} will have to process during inference in operation. It is on out-of-sample data that acceptable generalization behavior (and a corresponding guarantee) of the implemented \ac{mlm} can be reasonably expected. 

	\item \emph{In-MLMODD} $(\inmlmodd)$: Data that the implemented \ac{mlm} will have to process during inference in operation. In-MLMODD data contribute to the intended function(s) of the \ac{mlm}. We have: 
$\inmlmodd = \insample \cup \outofsample$ and  
$\insample \cap \outofsample = \emptyset$

	\item \emph{Out-of-MLMODD} $(\outmlmodd)$: Data not seen during \ac{mlm} learning that the implemented \ac{mlm} \emph{should not process} during inference in operation. Out-of-MLMODD data contributes to the intended function(s) of the \ac{mlc}, \eg specific processing to detect anomalies (see Section~\ref{s:odd-impact} for more details). We have: $\inmlmodd \cap \outmlmodd = \emptyset$.

	\item \emph{In-MLCODD} $(\inmlcodd)$: Data contributing to the intended function(s) of the \ac{mlc}. We have: $\inmlcodd = \inmlmodd \cup \outmlmodd$.

	\item \emph{Out-of-MLCODD} $(\outmlcodd)$: Data not seen during \ac{mlm} learning that the implemented \ac{mlc} \emph{should not process} during inference in operation. Out-of-MLCODD data contributes to the intended function(s) of the ML-based subsystem. We have: $\inmlcodd \cap \outmlcodd = \emptyset$.

\end{compactenum} 

\emph{Real Data in Operation} (and their associated statistical distributions), $\rdo$, can now be defined as the set of all data seen in operation: $\rdo \supseteq (\inmlcodd \setminus \inlier) \cup \outmlcodd$. 

The preceding concepts will serve as reference terms in forthcoming aviation industry specific guidance~\cite{as6983-draft}. Nevertheless, we believe they are generic enough to be applicable in other domains, although there are some differences, \eg our concept of \emph{Edge Case} data differs from what is considered in~\cite{ul4600std}.

\subsection{Illustrative Example (Aircraft Flight Envelope)}\label{s:application}

\newcommand{\alt}[1]{{$\mathtt{Alt:}$~\SI{#1}{\feet}}}
\newcommand{\mach}[1]{{$\mathtt{Mach}\;#1$}}

We now give an illustrative example of an aircraft flight envelope to concretize the preceding concepts. Informally, a flight envelope specifies the allowable combinations of two parameters---altitude ($\mathtt{Alt}$) and airspeed, given here as a Mach number ($\mathtt{Mach}$)---at which an aircraft design should function. 
Intuitively, this characterization of a flight envelope represents an \acf{od} of the aircraft system, and we refer to it, henceforth, as the \emph{system} \ac{od} (SOD). This is closely related to a \emph{functional} \ac{od} for the system which may include a specification of, for example, aircraft takeoff gross weights, the city pairs between which flight operations are intended, the routes (flight paths) that aircraft of a particular type design are expected to fly, the airports involved, the climb segments, and the landing approaches to be followed.

\begin{figure}[!htb]
	\centering
	\includegraphics[width=\textwidth]{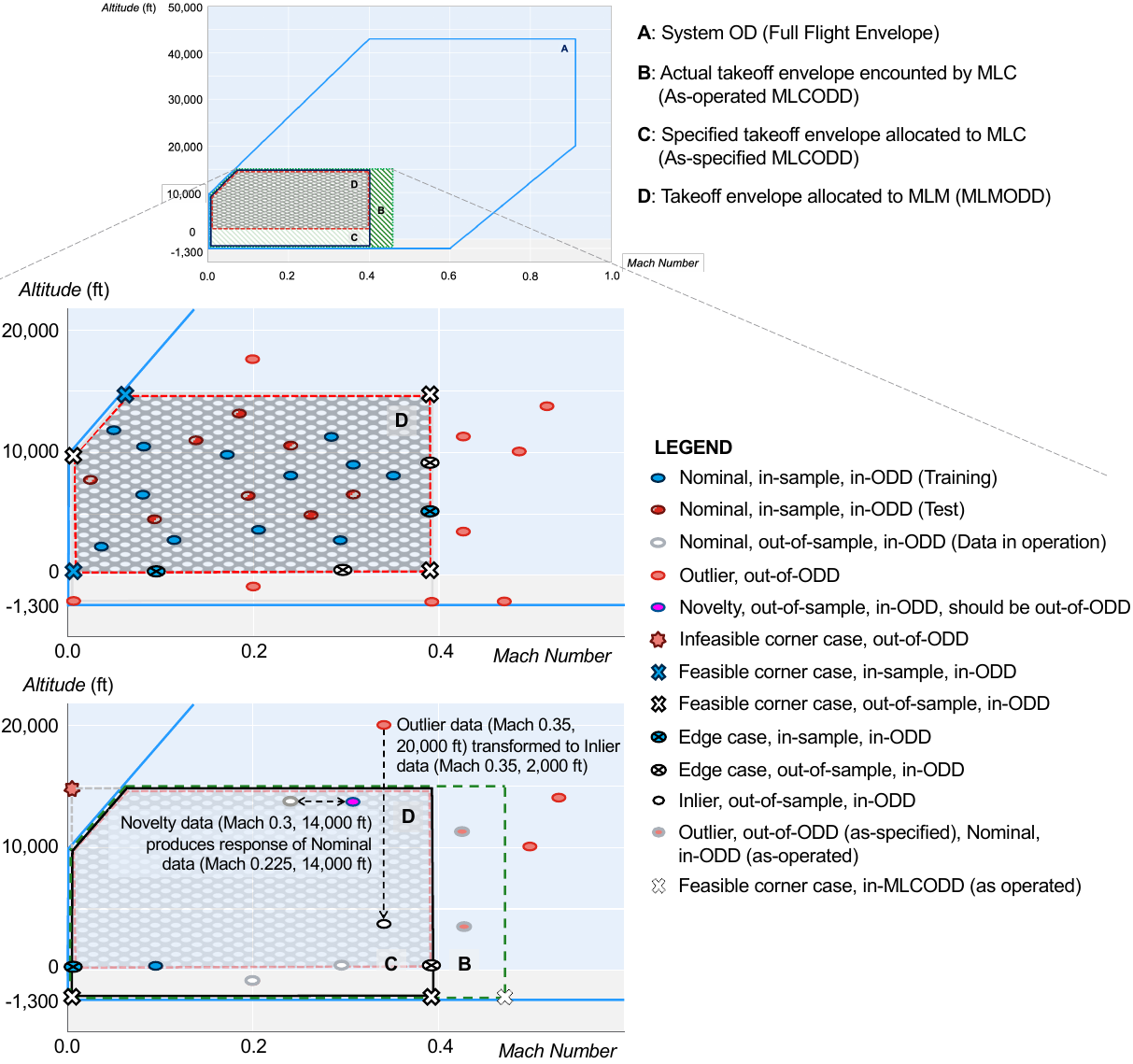}
	\caption{An example flight envelope (region A) representing an aircraft system \ac{od}, whose refinement and allocation to an \ac{mlc} and \ac{mlm} give, respectively, an \emph{As-operated} MLCODD (region B), containing an \emph{As-specified} MLCODD (region C), itself containing the MLMODD (region D), for the takeoff regime. The shapes representing the different \acp{odd} are practically congruent, but have been shown slightly offset here to differentiate each from the other. Zoomed-in views of the respective \acp{odd} highlight the different categories and kinds of data used to characterize them.}
	\label{f:odd-flight-envelope-example}
\end{figure}

Figure~\ref{f:odd-flight-envelope-example} presents a notional flight envelope covering all phases of flight (shown as the irregular hexagonal region A). $\mathtt{Mach}$ and $\mathtt{Alt}$ values within this SOD are allowed, and therefore they are expected to be encountered in operation. Values of those parameters outside the SOD are disallowed since operating outside the flight envelope is usually dangerous in most circumstances. 

Consider that a portion of this SOD is allocated to an ML-based subsystem to be used during the \emph{takeoff} flight phase. Its \ac{odd} (not shown in Figure~\ref{f:odd-flight-envelope-example}) is the takeoff regime at the bottom of the SOD which, in turn, we refine and allocate to an \ac{mlc} (contained by that ML-based subsystem). The resulting MLCODD parameters are: $0 \leq \mathtt{Mach} \leq 0.4$ and $\SI{-1300}{\feet} \leq \mathtt{Alt} \leq \SI{15000}{\feet}$. In Figure~\ref{f:odd-flight-envelope-example}, this \emph{As-specified} MLCODD is the irregular pentagon with the solid dark border (region C).

For this \ac{odd}, observe that the upper bound for the airspeed parameter is \mach{0.4}. However, aircraft with greater maximum takeoff weights, \eg cargo aircraft, can often exceed this bound during takeoff. Thus, there are two possibilities: either the design was to be restricted to non-cargo aircraft, or there is a missing requirement that would be discovered in operation with cargo aircraft. For the latter case, the \emph{as-operated} MLCODD would then have an increased upper bound on airspeed, \eg $0 \leq \mathtt{Mach} \leq 0.5$. In Figure~\ref{f:odd-flight-envelope-example}, this is the irregular pentagon (region B) that includes the earlier \emph{As-specified} MLCODD (region C).
Now, further consider that there is insufficient takeoff data for altitudes below sea-level to apply \ac{ml}. Hence, we restrict the \ac{mlm} to takeoff operations for $\mathtt{Alt} \geq$ \SI{0}{\feet}. Thus, the MLMODD is a sub-region of the MLCODD allocated to the \ac{mlm} contained in the \ac{mlc}. In Figure~\ref{f:odd-flight-envelope-example}, this is the irregular pentagon with the dashed border (region D), with the same range for $\mathtt{Mach}$ as its containing MLCODD, but with sea-level as the lower bound on $\mathtt{Alt}$. Figure~\ref{f:odd-flight-envelope-example} also zooms into these regions to give examples of the various categories and kinds of data described earlier. 

Data inside the MLMODD (and/or MLCODD) can be drawn from the \emph{Nominal}, \emph{Edge Case}, \emph{Feasible Corner Case}, \emph{Inlier}, and \emph{Novelty} data categories. The following observations are noteworthy: first, the \ac{mlm} must demonstrate generalization from \emph{Nominal}, \emph{In-Sample} training data to \emph{Nominal}, \emph{In-Sample} test data, as well as to \emph{Nominal}, \emph{Out-of-Sample} data, all of which are \emph{In-MLMODD}. Moreover, the \ac{mlm} must exhibit \emph{correct} behavior (\ie the behavior meets the allocated requirements) on \emph{Edge Case} as well as \emph{Feasible Corner Case} data. 	

Next, the preceding data categories are disjoint relative to a specific allocation. For example, \emph{Outlier} data for an \ac{mlm} cannot also be a \emph{Feasible Corner Case} for that \ac{mlm}, though it can be one for the containing \ac{mlc}. In Figure~\ref{f:odd-flight-envelope-example}, the data point (\mach{0.4}, \alt{-1300}) is one such example of an \emph{Outlier} for the MLMODD that is also an \emph{Out-of-Sample}, \emph{Feasible Corner Case} for the containing \emph{As-specified} MLCODD.
	
We associate data points with specific categories relative to an allocation. In Figure~\ref{f:odd-flight-envelope-example} for instance, the data point at (\mach{0.1}, \alt{0}) is an \emph{Edge Case} for the MLMODD, but is \emph{Nominal} data for the containing \emph{As-specified} MLCODD. Similarly, each of the data points at (\mach{0}, \alt{0}), and (\mach{0.4}, \alt{0}) is a \emph{Corner Case} from an MLMODD perspective but an \emph{Edge Case} for the MLCODD.

Likewise, we can have \emph{Outlier} data to the MLMODD that are within the MLCODD. In Figure~\ref{f:odd-flight-envelope-example}, examples of this case comprise any data point in the region of the \emph{As-specified} MLCODD not included in the MLMODD, \ie in the region defined by $0 \leq \mathtt{Mach} \leq 0.4$, and $\SI{-1300}{\feet} \geq \mathtt{Alt} > \SI{0}{\feet}$. As shown, such points are \emph{Outlier} data for the MLMODD, but can be \emph{Nominal}, \emph{Edge Case} or \emph{Feasible Corner Case} data for the MLCODD. In the same way, points in the rectangular region of the takeoff envelope between $0.4 < \mathtt{Mach} \leq 0.5$ and $\SI{0}{\feet} \leq \mathtt{Alt} \leq \SI{15000}{\feet}$ are \emph{Outlier} data to both the MLMODD, and the \emph{As-specified} MLCODD, but are within the \emph{As-operated} MLCODD. For example, the data point at (\mach{0.5}, \alt{-1300}) is a \emph{Feasible Corner Case} for the \emph{As-operated} MLCODD.

Recall that \emph{Infeasible Corner Case} data are a special case of \emph{Outlier} data that may not be reasonably encountered in operation, where two or more \ac{odd} parameters simultaneously take the extreme values admissible for that \ac{odd}. Figure~\ref{f:odd-flight-envelope-example} (bottom right) shows one such example: the corner case at (\mach{0.0}, \alt{15000}) is infeasible for both the MLMODD and MLCODD because no airport runways exist at  \SI{15000}{\feet} altitude.

\emph{Inlier} data are within the MLCODD and/or MLMODD due to errors in data processing, scaling, normalization, and usage of incorrect units. In Figure~\ref{f:odd-flight-envelope-example}, the \emph{Inlier} data point at (\mach{0.35}, \alt{2000}) is the result of a data preparation and scaling error of the \emph{Outlier} data point at (\mach{0.35}, \alt{20000}). The result of processing such \emph{Inlier} data is an incorrect response from the \ac{mlm}, for example a flight control parameter value appropriate for the outlier data point is incorrectly produced at a lower altitude within the takeoff envelope. 

\emph{Novelty} data are within the MLMODD (and thus, also within the MLCODD), but are, in fact, data that should have been Out-of-MLMODD (or MLCODD). \emph{Novelty} data are not excluded from the MLMODD due to an insufficiency in the number and variety of parameters used to specify the MLMODD. 

In Figure~\ref{f:odd-flight-envelope-example}, the data point (\mach{0.3}, \alt{14000}) is \emph{Novelty} data producing a response appropriate for the \emph{Nominal} data point at (\mach{0.225}, \alt{14000}). 
This occurs because the SOD and, in turn, the MLCODD and MLMODD have been specified using only two parameters (altitude and airspeed), either ignoring the effect of additional parameters such as air temperature, or implicitly assuming that the operations occur in the same environment as that in which the in-sample data were collected. In this example, operating in warmer air temperatures results in a lower Mach number, due to which the \ac{mlm} receives an input that is invalid for the operating context, but is nonetheless \emph{Nominal}.  

In general, discovering data from the \emph{Inlier}, \emph{Novelty}, and \emph{Outlier} categories that should be part of the required (or intended) MLCODD or MLMODD, occurs either during testing, during validation of the relevant \acp{odd}, or from analysis of the data gathered from in-service experience. That usually results in re-defining the respective \acp{odd}, \eg by expanding its dimensions by including additional parameters, or modifying the admissible range of existing parameter values.

\section{Support for System-level Analysis} \label{s:odd-impact}

The combination of the category and kind of real data in operation, $\rdo$,  facilitates \emph{partitioning} an MLMODD (and equivalently, an MLCODD) at a higher level than, say, partitioning by equivalence classes of inputs.\footnote{In fact, we can combine those two ways of partitioning an \ac{odd}, \eg by selecting an equivalence class of inputs within a \emph{nominal}, \emph{out-of-sample}, and \emph{In-MLMODD} partition.}
Then, from a safety standpoint for example, for each such partition we can analyze the contribution of the \ac{mlm} (or the corresponding \ac{mlc}) to system hazards in terms of the effects produced in response to inputs drawn from that partition.
Examples of such effects include: an underperformance of the \ac{mlm}; a hazardous failure condition; \ac{mlm} or \ac{mlc} malfunction; or, more generally, \ac{mlm} and \ac{mlc} failure modes and \emph{hazard contribution modes}~\cite{dps-safeai2020}.

\begin{figure}[!htb]
	\centering
	\includegraphics[width=\textwidth]{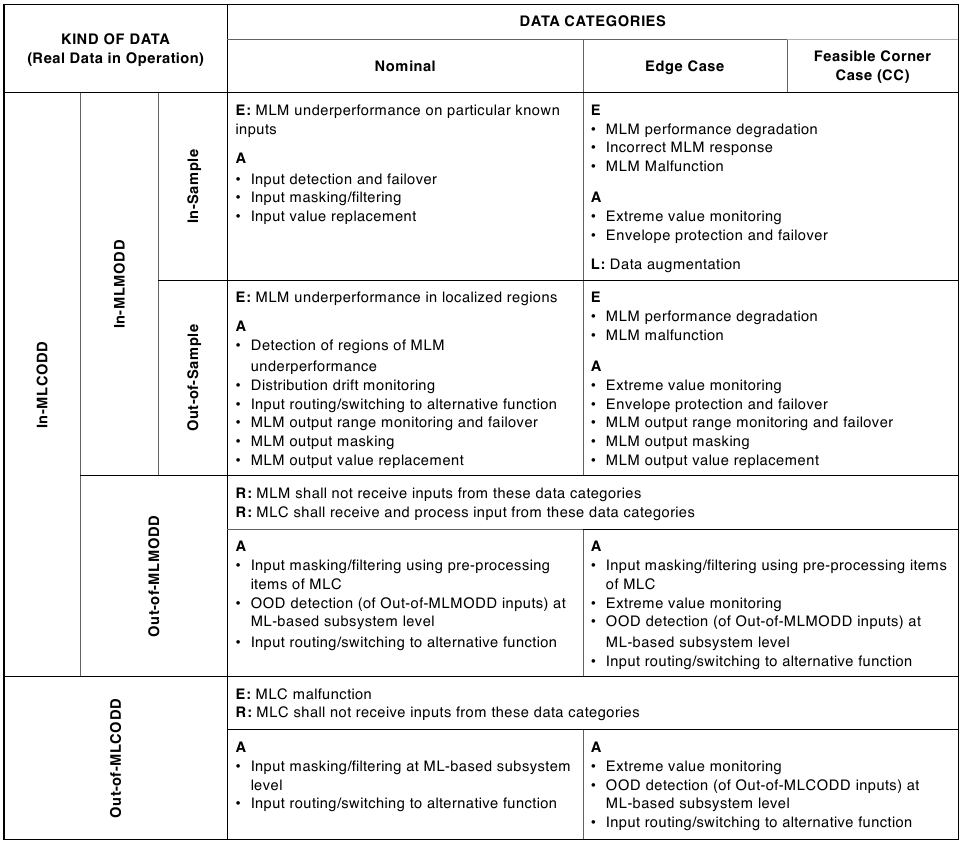}	
	\caption{Assessing the impact of an \ac{odd} on an \ac{mlm} and the corresponding \ac{mlc} in relation to the partitions induced by the categories and kinds of real data in operation (specifically the \emph{Nominal}, \emph{Edge Case}, and \emph{Feasible Corner Case} data categories) described in terms of the potential \emph{effects} ({\bf \textsf{E}}) of the data, the \emph{requirements} ({\bf \textsf{R}}) induced, the \emph{learning assurance} ({\bf \textsf{L}}) processes that may be needed, and candidate \emph{architectural} ({\bf \textsf{A}}) options for mitigation.}
	\label{f:odd-arch-non-anomaly}
\end{figure}

\begin{figure}[!htb]
	\centering
	\includegraphics[width=\textwidth]{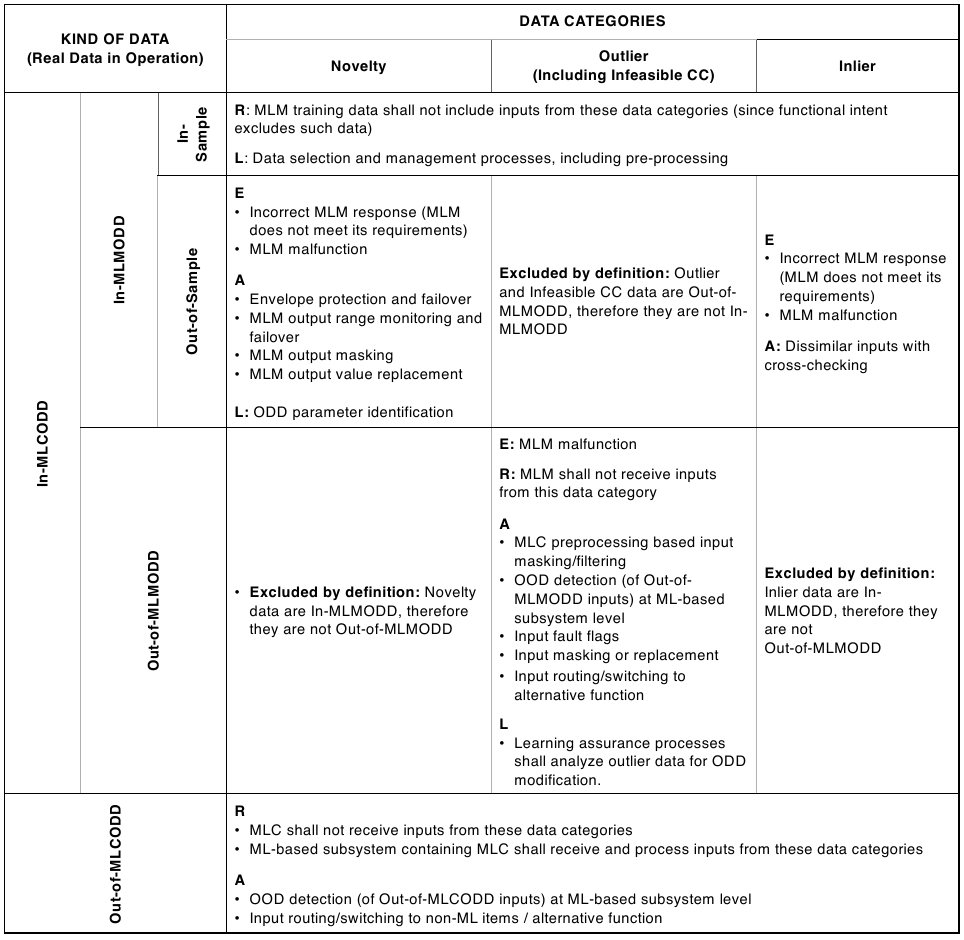}	
	\caption{Assessing the impact of \acp{odd} characterized by \emph{Anomaly} data, \ie \emph{Novelty}, \emph{Outlier} (including \emph{Infeasible Corner Case}), and \emph{Inlier} data categories, similar to the assessment in Figure~\ref{f:odd-arch-non-anomaly}.}
	\label{f:odd-arch-anomaly}
\end{figure}

Subsequently, we can establish the (high-level) requirements that an \ac{mlm} and its containing \ac{mlc} should fulfill. These can include, for instance, restrictions on \ac{mlm} behavior, constraints on data processing, limitations of use, as well as requirements necessary to manage the safety impact of the identified effects. The latter, in turn, also informs the selection of the mitigation measures appropriate for sufficient safety assurance. 
Such mitigations include the application of learning assurance processes (at the \ac{mlm} layer), architectural mechanisms (at the \ac{mlc}, ML-based subsystem, and system layers), as well as traditional development assurance processes as appropriate. 

The tables given in Figure~\ref{f:odd-arch-non-anomaly} and Figure~\ref{f:odd-arch-anomaly} illustrate how we can use the partitions of an \ac{odd} to analyze the impact on an \ac{mlc} and \ac{mlm}: the row and column labels for a cell in the table correspond to the kinds and categories of data, respectively, and their combination is the partition of $\rdo$ we analyze. 
The content of a cell describes the results of a particular analysis for that partition, \ie the effects of encountering data from that partition, and the considerations that emerge on the requirements, architectural mitigations, and on learning assurance. When the analysis is common to multiple partitions, we show this in a cell that spans multiple columns. 
Note that these kinds of analyses can be applied to any \ac{ml}-based subsystem, \ac{mlc}, or \ac{mlm}, and is agnostic to their allocated function. Also note that Figure~\ref{f:odd-arch-non-anomaly} and Figure~\ref{f:odd-arch-anomaly} are mainly examples, hence they are not comprehensive or complete. Thus, some effects (and the corresponding architectural options) can be common to the different partitions. 

For brevity, here we highlight some specific example options from a combination of analyses. In practice, however, each analysis would be separately undertaken since the identified learning assurance techniques only apply during design, whereas the identified architectural options are primarily relevant in use. 

Figure~\ref{f:odd-arch-non-anomaly} shows an analysis from a safety standpoint: the potential effects of the \ac{odd} partition characterized by \emph{In-MLMODD}, \emph{In-Sample}, \emph{Nominal} data include \ac{mlm} underperformance on specific inputs (as observed during training and testing). In some applications, the exact inputs from that partition may also be encountered in operation. Thus, architectural mitigations for such data can include monitoring to detect those specific inputs, together with input value replacement, masking, or filtering, and/or failover. 

Similarly, other partitions of the \ac{odd} can be characterized by 
\emph{In-MLMODD}, \emph{Out-of-sample}, \emph{Edge Case} (or \emph{Feasible Corner Case}) data. Figure~\ref{f:odd-arch-non-anomaly} shows these combined into a single partition since the high-level effects (such as \ac{mlm} malfunction), as well as the corresponding architectural mitigations (\eg extreme value monitoring, or envelope protection and failover) are similar for each. However, we note that for particular applications involving a specific \ac{mlm}, the individual effects (and therefore the necessary architectural mitigations) from \emph{Edge Case} inputs are likely to differ from those resulting from \emph{Feasible Corner Case} inputs.

Likewise, a common requirement is induced by the \ac{odd} partition(s) formed by (each of) the \emph{In-MLCODD}, \emph{Out-of-MLMODD}, \emph{Nominal} data (and \emph{Edge Case}, or \emph{Feasible Corner Case} data respectively). For example, an \ac{mlm} shall not receive and process input data drawn from those partitions of the \ac{odd}. Consequently, the architectural options available are also largely similar, although extreme value monitoring mainly applies to \emph{Edge Case} and \emph{Feasible Corner Case} data, rather than to \emph{Nominal} data.
Additionally, note that for \emph{Out-of-MLCODD} kind of data, there is no distinction between \emph{Nominal}, \emph{Edge Case} and \emph{Corner Case} data from an \ac{mlc} standpoint. However, those categories are distinct from the perspective of the \ac{od} allocated to the containing ML-based subsystem, which induces distinct architectural mitigations as shown.

Figure~\ref{f:odd-arch-anomaly} shows a similar analysis from a system development standpoint for the \emph{Novelty}, \emph{Outlier} (including \emph{Infeasible Corner Case}), and \emph{Inlier} data categories. Such data are not part of the functional intent, and therefore a requirement on \ac{mlm} development is to exclude such data for model training. As such, the learning assurance process must include data selection and management activities to assure that the training data indeed excludes inputs drawn from those data categories to preclude an \ac{mlm} from producing responses that are inconsistent with the functional intent. 
\optional{Here, we note that it is possible to include \emph{Anomaly} data into model training if the functional intent is to manage those within the model itself. In that case, \emph{Anomaly} data would rather be categorized as \emph{Nominal} data, but further partitioned as one or more special equivalence classes within the \emph{Nominal} data category.}

\optional{It is important to test the \ac{mlc} robustness against \emph{Outlier} data. In particular exercising \emph{Corner Case} data---especially \emph{Infeasible Corner Case} data---is standard robustness testing practice.}

The partitions of the \ac{odd} characterized by \emph{In-MLMODD}, \emph{Out-of-sample}, \emph{Novelty} (and likewise \emph{In-MLMODD}, \emph{Out-of-Sample}, \emph{Inlier}) data need special attention: specifically, \emph{Novelty} data may not be detectable through operational monitoring. Indeed, if such data could be detected at runtime, the relevant features would then have been included in the set of MLMODD parameters, rendering such data \emph{Nominal} rather than \emph{Novelty}. \emph{Inlier} data are also similarly difficult to detect in operation. 
To mitigate \ac{mlm} failure conditions resulting from the former, learning assurance activities are particularly important, especially those facilitating a rigorous and comprehensive identification of MLMODD parameters and features. 

In some circumstances, it may be possible to detect and recover from the \emph{effects} of \emph{Novelty} data if the responses produced result in a range violation. For those situations a range of output monitoring, masking, replacement, and failover mechanisms offer an architectural solution to risk mitigation. To mitigate the effects of \emph{Out-of-Sample}, \emph{In-MLMODD}, \emph{Inlier} data, dissimilar and/or independent inputs with cross-checking is a candidate architectural pattern.

An \ac{mlm} cannot receive \emph{In-MLMODD}, \emph{Out-of-Sample},
\emph{Outlier} data since those are, by definition, \emph{Out-of-MLMODD}. However, in a similar vein to \emph{Novelty} and \emph{Inlier} data, the \ac{odd} partition characterized by \emph{In-MLCODD}, \emph{Out-of-MLMODD}, \emph{Outlier} data also needs particular attention: 
as we saw earlier (Section~\ref{s:application}, Figure~\ref{f:odd-flight-envelope-example}), it is possible to encounter \emph{Outlier} data that \emph{ought to have been included} in the MLCODD---and by allocation, also in the MLMODD---but were not. This situation can occur due to an error in the requirements, a deficiency in the data collection process, or a lack of knowledge (epistemic uncertainty). This induces a learning assurance feedback step (see Figure~\ref{f:odd-arch-anomaly}) to analyze \emph{Outlier} data to validate and potentially update both the MLMODD and the MLCODD from in-service experience.

\section{Conclusions and Future Work}\label{s:concluding-remarks}

We have clarified the dimensions along which the parameters that define an \ac{odd} for an \ac{ml}-based aeronautical product can be captured, whilst identifying the categories and kinds of data that can be encountered in operation. We have concretized the underlying concepts using an aircraft flight envelope example considering its allocation to an \acf{mlm} for the takeoff regime. 
Our data-centric \ac{odd} characterization gives a useful framework to identify and organize system development, safety, and assurance activities, which we have illustrated through examples of some high-level effects of the data both on the \ac{mlm} and its containing \acf{mlc}, along with the architectural options available for mitigation. 


The work described here has emerged from an ongoing, aviation industry-led, consensus based effort. As such, validating the relevance, applicability, and utility of the underlying concepts and approach largely relies on a committee consensus agreement and, eventually, regulatory endorsement. 
To that end, aviation industry practitioners are applying the approach to a variety of real-world applications such as airborne collision avoidance~\cite{acas-xu-hybrid}, safe flight termination\footnote{See: \href{https://safeterm.eu/}{https://safeterm.eu/}}, and time-based separation of transport aircraft in terminal approaches~\cite{eurocontrol-coast-tbs}. These use cases corroborate our earlier assertion (see Section~\ref{s:odd-impact}) that the work in this paper is sufficiently generic to be applicable to \ac{ml}-based aeronautical products used both in airborne systems, and for air traffic management/navigation services.
%
%
As a key avenue of future work, we are committed to take the lessons learned from those validation efforts---of the successes, insights, and possible gaps---to refine and further mature our approach. A related, crucial aspect of our future effort is to leverage the concepts and approach presented here to define a rigorous process for MLCODD development and validation, and MLCODD coverage verification (to be elaborated in a forthcoming paper). Such a process does not yet exist in the prevailing aviation standards and guidance on recommended practices. Thus, it will represent a concrete extension to the state-of-the-practice. 
\optional{A core challenge therein is defining the process such that it is fully consistent both with existing aviation systems engineering practices, \eg~\cite{SAE-ARP-4761}, \cite{SAE-ARP-4754A}, as well as an \ac{ml} Development Lifecycle~\cite{mldl-sae-journal} that is also being developed as part of the same standardization effort motivating the work in this paper. In particular, special attention is required for the interfaces with
\begin{inparaenum}[(i)]
	\item system requirements that include system \ac{od} information, 
	\item the data management processes, and
	\item the \ac{mlm} design process.
\end{inparaenum}}	

Our data-centric \ac{odd} characterization (Section~\ref{s:concepts}), though rigorous, would benefit from \optional{a formal definition of the different \acp{odd} and their allocations to different system layers, as well as}a formalization of the identified categories and kinds of data, and their interrelations. This could facilitate assessing whether certain desirable properties hold, \eg that the data categories \emph{cover} an \ac{odd}\optional{---and, via the allocation through the system layers, also the \ac{od}---} in some formally defined sense, and that they are internally \emph{complete}. 
This paper has mainly considered singleton \acp{mlm} and \acp{mlc}. We intend to extend our approach to the situations of multiple \acp{mlc} within a single ML-based subsystem, and multiple sub-\acp{mlm} within a single \ac{mlc}. These cases have interesting safety and architectural implications \optional{owing to overlapping \ac{mlm}\acp{odd}, transition discontinuities when switching between those \acp{odd}, and the effects of the various data categories on the overlapping portions of different \acp{odd}. Thus}from which we expect to gain a deeper insight into hazardous behavior emerging from the interactions of multiple \acp{mlm} and \acp{mlc}. 
In a similar vein, the support for system-level analysis (Section~\ref{s:odd-impact}) can be further elaborated towards a more comprehensive and complete description of the potential effects of real data encountered in operation, together with the requirements induced, architectural options available for mitigation, and the learning assurance activities necessary. \optional{For example, we can determine the criticality of the identified effects by associating those effects with the severity of the worst-case consequence to which they contribute at the system level. That, in turn, establishes the level of assurance considered sufficient which can then aid in selecting a suitable architectural mitigation amongst the available options. A related area for future work is to elaborate the safety assurance argument that a data-centric \ac{odd} characterization implicitly embeds.}

\optional{The scope of \ac{ml} that has underpinned this work is primarily supervised, offline \ac{ml} with (deep) neural networks. A further topic for future work \optional{aligned with the validation effort mentioned earlier,}is to examine the \optional{suitability and} applicability \optional{of our approach} to other \ac{ml} schemes (\eg reinforcement learning)\optional{, aided by comprehensive examples covering both low- and high-dimensional aeronautical applications}.}

This paper has given a new data-centric characterization for \acp{odd} that is not an extension, enhancement, or tailoring of prior automotive domain \ac{odd} concepts. A related avenue of future work is to compare and contrast our \ac{odd} concept and principles with those of other safety-critical domains such as automotive, healthcare, rail, and space. We remain cautiously optimistic that our work is sufficiently general to be adopted, extended, and applied in those domains by the associated subject-matter experts.

\subsubsection*{Acknowledgments} 

We thank the members of the \ac{odd} working group of the joint EUROCAE WG-114 and SAE G-34 committees who contributed to the discussions that shaped the concepts and approach in this paper. We are additionally grateful to the anonymous reviewers whose comments aided us in improving the paper.

%
%

\begin{thebibliography}{10}
\providecommand{\url}[1]{\texttt{#1}}
\providecommand{\urlprefix}{URL }
\providecommand{\doi}[1]{https://doi.org/#1}

\bibitem{bsi-odd-pas}
{BSI Standards Ltd.}: {Operational Design Domain (ODD) Taxonomy for an
  Automated Driving System (ADS) -- Specification}. BSI PAS 1883:2020 (August
  2020)


\bibitem{acas-xu-hybrid}
Damour, M., et al.:
  {Towards Certification of a Reduced Footprint ACAS-Xu System: A Hybrid
  ML-Based Solution}. In: Habli, I., Sujan, M., Bitsch, F. (eds.) Computer
  Safety, Reliability, and Security. pp. 34--48. Springer, Cham (2021)


\bibitem{dps-safeai2020}
Denney, E., Pai, G., Smith, C.: {Hazard Contribution Modes of Machine Learning
  Components}. In: Espinoza, H., et al. (eds.) {Proceedings of the AAAI Workshop on Artificial Intelligence Safety (SafeAI)}. AAAI, CEUR Workshop Proceedings (2020)


\bibitem{eurocontrol-coast-tbs}
{EUROCONTROL}: {COAST (Calibration of Optimised Approach Spacing
  Tool) with Use of Machine Learning Models}. White Paper~V1.1. (April 2021)


\bibitem{easa-concept-paper}
{EASA}: {First Usable Guidance for Level 1 Machine Learning Applications}. 
EASA Concept Paper Issue 01 (December 2021)


\bibitem{SAE-AIR-6988}
{G-34, Artificial Intelligence in Aviation Committee}: {AIR 6988, Artificial
  Intelligence in Aeronautical Systems: Statement of Concerns}. {SAE
  International} (April 2021)

\bibitem{odd-definition-language}
Irvine, P., Zhang, X., Khastgir, S., Schwalb, E., Jennings, P.: {A Two-Level
  Abstraction ODD Definition Language: Part I}. In: 2021 IEEE International
  Conference on Systems, Man, and Cybernetics (SMC). pp. 2614--2621 (2021).

\bibitem{iso-pas-8800}
{ISO/TC 22/SC 32}: {Road Vehicles - Safety and Artificial Intelligence}.
  ISO/AWI PAS 8800 (Under development) (2021)

\bibitem{iso-dis-34503}
{ISO/TC 22/SC 33}: {Road vehicles - Test Scenarios for Automated Driving
  Systems - Taxonomy for Operational Design Domain}. ISO/DIS 34503 - Draft
  International Standard (2023)


\bibitem{mldl-sae-journal}
Kaakai, F., Adibhatla, S., et al.: 
{Toward a Machine Learning Development Lifecycle for Product Certification 
and Approval in Aviation}. SAE Intl. Journal of Aerospace 
\textbf{15} (2022)

\bibitem{micro-odd}
Koopman, P., Osyk, B., Weast, J.: Autonomous Vehicles meet the Physical World:
  RSS, Variability, Uncertainty, and Proving Safety. In: Romanovsky, A.,
  Troubitsyna, E., Bitsch, F. (eds.) Computer Safety, Reliability, and
  Security. pp. 245--253. Springer, Cham (2019)

\bibitem{nhtsa-safety-vision}
{NHTSA, US Department of Transportation}: {Automated Driving: A Vision for
  Safety}. Report No. DOT HS 812 442 (September 2017)

\bibitem{SAE-ARP-4761}
{S-18, Aircraft And System Development And Safety Assessment Committee}: {ARP
  4761, Guidelines and Methods for Conducting the Safety Assessment Process on
  Civil Airborne Systems and Equipment}. {SAE International} (Dec 1996)

\bibitem{SAE-ARP-4754A}
{S-18, Aircraft And System Development And Safety Assessment Committee}: {ARP
  4754A, Guidelines for Development of Civil Aircraft and Systems}. {SAE
  International} (Dec 2010)

\bibitem{as6983-draft}
{SAE G-34 Committee for AI in Aviation and EUROCAE WG-114 for AI}: {Process
  Standard for Development and Certification/Approval of Aeronautical
  Safety-Related Products Implementing AI}. AS 6983 Draft Standard {Work In
  Progress} (February 2023)

\bibitem{sae-j3016}
{SAE International}: {Taxonomy and Definitions for Terms Related to Driving
  Automation Systems for On-Road Motor Vehicles}. {Surface Vehicle Recommended
  Practice J3016} (2018)

\bibitem{ads-tests-scenarios}
Thorn, E., Kimmel, S., Chaka, M.: {A Framework for Automated Driving System
  Testable Cases and Scenarios}. Report No. DOT HS 812 623, National Highway
  Traffic Safety Administration (September 2018)

\bibitem{ul4600std}
{Underwriter Laboratories Inc.}: {ANSI/UL 4600 Standard for Safety for the
  Evaluation of Autonomous Products} (April 2020)

\end{thebibliography}

\end{document}